# Tracing the history of an unusual compound presolar grain from progenitor star to asteroid parent body host

S. A. Singerling[1*], L. R. Nittler[2,5], J. Barosch[2], E. Dobrică[3], A. J. Brearley[4], and R. M. Stroud[1,5]

[1]U.S. Naval Research Laboratory, Code 6366, Washington, DC 20375, USA
[2]Carnegie Institution of Washington, Washington, DC 20015, USA
[3]University of Hawai'i at Mānoa, Honolulu, HI 96822, USA
[4]University of New Mexico, Albuquerque, NM 87131, USA
[5]Arizona State University, Tempe, AZ 85287, USA

*NRC Postdoc; corresponding author current institution: Virginia Polytechnic Institute and State University, Blacksburg, VA 24061, USA; email: ssheryl@vt.edu

**Abstract**

We conducted a transmission electron microscopy (TEM) study of an unusual oxide-silicate composite presolar grain (F2-8) from the unequilibrated ordinary chondrite Semarkona (LL3.00). The presolar composite grain is relatively large (>1 µm), has an amoeboidal shape, and contains Mg-rich olivine (forsterite), Mg-Al spinel, and Ca-rich pyroxene. The shape and phase assemblage are reminiscent of amoeboid-olivine-aggregates (AOAs) and add to the growing number of TEM observations of presolar refractory inclusion-like (CAIs and AOAs) grains. In addition to the dominant components, F2-8 also contains multiple subgrains, including an alabandite-oldhamite composite grain within the olivine and several magnetite subgrains within the Mg-Al spinel. We argue that the olivine, Mg-Al spinel, and alabandite-oldhamite formed by equilibrium condensation, whereas the Ca-rich pyroxene formed by non-equilibrium condensation, all in an M-type AGB star envelope. On the other hand, the magnetite subgrains are likely the result of aqueous alteration on the Semarkona asteroidal parent body. Additional evidence of secondary processing includes Fe-enrichment in the Mg-Al spinel and olivine, elevated Al contents in the olivine, and beam sensitivity and a modulated structure for the olivine.

Compound presolar grains, in particular oxide-silicate AOA-like grains such as F2-8, record condensation conditions over a wide range of temperatures. Additionally, the presence of several different presolar phases in a composite grain can impart information on the relative rates and effects of post-condensation processing in a range of environments, including the interstellar medium, solar nebula, and the host asteroid parent body. For example, the olivine and spinel in F2-8 show evidence of fluid infiltration, but each component reacted in different ways and to different extents. The TEM observations of F2-8 provide insights across the lifetime of the grain from its formation by condensation in an M-type AGB star envelope, its transit through the interstellar medium, and aqueous alteration during its residence on Semarkona's asteroidal parent body.

# 1. INTRODUCTION

Presolar grains are some of the original building blocks of our Solar System, predating the formation of the solar nebula by up to several billion years (Heck et al., 2020). They are identified by their extremely anomalous isotopic compositions, compared to materials that formed in the Solar System. Presolar grains come in a variety of phases, including carbides, graphite, diamond, oxides, and silicates (Lodders and Amari, 2005; Zinner, 2014; Nittler and Ciesla, 2016). Their sizes range from the nm to μm scale. Within presolar oxide and silicate phases alone, which are most pertinent to the current study, previously identified oxides include corundum ($Al_2O_3$), Mg-Al spinel ($MgAl_2O_4$), and less commonly titanium dioxide ($TiO_2$), chromite ($(Mg,Fe)Cr_2O_4$), magnetite ($Fe^{2+}Fe_2^{3+}O_4$), and hibonite ($CaAl_{12}O_{19}$), and previously observed silicates include olivine ($(Mg,Fe)_2SiO_4$), pyroxene ($Ca_{0-1}(Mg,Fe)_{1-2}Si_2O_6$), non-stoichiometric materials, and less commonly silica ($SiO_2$) and perovskite-structured $MgSiO_3$ (Lodders and Amari, 2005; Zinner, 2014; Floss and Haenecour, 2016; Nittler and Ciesla, 2016). Composite presolar grains consisting of multiple mineral phases have also been identified, some of which contain refractory phases and are reminiscent of some of the earliest formed solids in the Solar System, calcium-aluminum-rich inclusions (CAIs) and amoeboid olivine aggregates (AOAs) (Vollmer et al., 2009b; Nguyen et al., 2010; Floss and Stadermann, 2012; Vollmer et al., 2013; Nguyen et al., 2014; Leitner et al., 2018; Nittler et al., 2018). Unaltered AOAs are mostly composed of forsterite and CAIs, which themselves consist of spinel, pyroxene, and anorthite (e.g., Krot et al. 2019).

Presolar grains are divided into groups based on their isotopic compositions, which provide insights into their formation. For oxides and silicates, the main groups (1–4) are based on O isotopes (Nittler et al., 1997). The isotopic compositions of presolar grains are inconsistent with formation by any known mechanisms in our Solar System, but instead, are the products of condensation of the nucleosynthetic products of different circumstellar environments. The majority of oxide and silicate presolar grains have isotopic compositions consistent with origins in the outflows of evolved low mass stars (<8 $M_☉$) in their asymptotic giant branch (AGB) phases. Dust formation occurs around AGB stars when their atmospheres cool enough that solids condense from atoms, ions, and molecules dredged up from their interiors, and the dust is carried to the interstellar medium coupled to gaseous outflows from the pulsating stellar core.

Low abundances of presolar circumstellar stardust grains are preserved in the matrices of primitive meteorites (ppb to a few 100 ppm, matrix normalized; Lodders and Amari, 2005; Zinner, 2014; Nittler and Ciesla, 2016). Ordinary chondrites are one group of primitive meteorites which have escaped complete melting, and so retain pristine features, including presolar grains (e.g., Alexander et al., 1990; Scott and Krot, 2014). The LL3.0 ordinary chondrite Semarkona is especially pristine, only exhibiting effects from interaction with fluids (e.g., Hutchison et al., 1987; Alexander et al., 1989; Krot et al., 1997; Grossman and Brearley, 2005). Additionally, amorphous silicate-rich regions of relatively pristine matrix material, which largely escaped aqueous alteration, have been identified (Dobrică and Brearley, 2020). In general, parent body processing decreases the abundance of presolar grains by destroying them or diluting their anomalous isotopic compositions and can increase the Fe contents of presolar silicates and oxides (Floss and Hanenecour, 2016). A NanoSIMS (nanoscale secondary ion mass spectrometry) search of these regions identified abundant presolar grains, consistent with minimal secondary processing (Barosch et al., 2022). In a companion paper, we reported and discussed transmission electron microscopy (TEM) data for several presolar SiC, oxide, and silicate grains studied within Semarkona (Singerling et al., 2022).

Here, we focus on a presolar composite grain (F2-8) from Semarkona. The grain is unusually large (>1 µm) and has a morphology and phase assemblage reminiscent of AOAs. We employed TEM to investigate the structural and chemical composition characteristics of the grain in an effort to determine what processes and conditions were responsible for the grain's formation in the atmosphere of its progenitor AGB star, as well as any secondary processing it was exposed to in the interstellar medium (ISM), in the solar nebula, and/or on its asteroidal parent body. Compound presolar grains, in particular oxide-silicate AOA-like grains, record condensation conditions over a wide range of temperatures. Additionally, the presence of several different presolar phases in a composite grain can impart information on the relative rates and effects of post-condensation processing in a range of environments from the interstellar medium, solar nebula, and the host asteroid parent body.

## 2. METHODS

### 2.1 NanoSIMS

The presolar grain F2-8 was detected during an automated NanoSIMS isotopic imaging search using the Cameca NanoSIMS 50L at the Carnegie Institution (Barosch et al., 2022) on the same Semarkona thin section (UNM 102) studied by Dobrică and Brearley (2020). A $Cs^+$ primary ion beam (0.5 pA; 100–120 nm beam size) was used to map $^{12-13}C^-$, $^{16-18}O^-$, $^{28}Si^-$, and $^{27}Al^{16}O^-$ isotopes and identify presolar grains. More details about analytical conditions and data reduction can be found in Barosch et al. (2022). Following its identification, we used a Hyperion RF plasma source (Oregon Physics, LLC) to re-measure grain F2-8 for its Mg, Al, and Si isotopic compositions. A 2.4 pA $O^-$ beam of 120 nm size was used with simultaneous collection of $^{24-26}Mg^+$, $^{27}Al^+$, and $^{28-30}Si^+$ secondary ions. Isotope ratios were normalized to the surrounding matrix.

### 2.2 Focused Ion Beam Sample Preparation

We used the focused ion beam (FIB) sample preparation technique to prepare the presolar grain and surrounding matrix material for TEM analyses. We used a FEI Helios G3 FIB-SEM at the U.S. Naval Research Laboratory to perform in situ liftout of a FIB section containing presolar composite grain F2-8. In order to identify the locations of presolar grains for liftout, we aligned SIMS isotope images and SEM-FIB secondary electron images as image overlays, and deposited C and Pt fiducial markers on the grain centers and directions of thinning. FIB section preparation involved: 1) depositing fiducial markers on the presolar grains followed by a protective C or Pt strap using a FEI Multichem Gas Injection System; 2) cutting a lamella of the grain from the surrounding matrix with a $Ga^+$ ion beam; 3) extracting the lamella using a FEI EasyLift NanoManipulator System; 4) welding the lamella onto a TEM Cu half grid with Pt deposition; and 5) thinning the lamella to electron transparency (i.e., 100 nm) using progressively smaller ion beam currents and accelerating voltages. The working distance was 4 mm. Imaging using the electron beam was done at 5 kV and 1.6 nA. Conditions using the ion beam varied—30 kV and 25 or 40 pA for imaging, 30 kV and 0.79 or 2.5 nA for milling, 30 kV and 40 or 80 pA for deposition, 30 kV and 0.23 nA for initial thinning, and 16 kV and 50 down to 23 pA for final thinning.

### 2.3 Transmission Electron Microscopy

We analyzed the presolar grain for structural and elemental compositional information on two TEMs—a JEOL 2200FS TEM and a Nion UltraSTEM-200X at the U.S. Naval Research Laboratory, both operating at 200 kV with double tilt holders. The JEOL was primarily used for microstructural work, whereas the Nion was used for compositional work, including energy dispersive X-ray spectroscopy (EDS) and electron energy loss spectroscopy (EELS). On the JEOL, a Gatan OneView® camera was used to collect bright field (BF) and high resolution (HR) images as well as selected area electron diffraction (SAED) patterns. Image magnification calibrations were performed with a Mult-I-Cal™ Si-Ge lattice standard, whereas the diffraction camera lengths were calibrated with a polycrystalline Al diffraction standard. In some instances, the small sizes of the different regions of interest precluded the use of SAED patterns for meaningful microstructural information. In those cases, we instead performed Fast Fourier Transforms (FFTs) on representative regions within HRTEM images in order to identify phases as well as get more detailed information on the crystallinity.

For phase identification, we measured the d-spacings of diffraction spots and, in the case of SAED patterns also angles, and compared those to the literature values of phases with appropriate compositions and that have been observed in presolar materials (Table S1). The diffraction data were checked against the following phases: forsterite, fayalite, enstatite, clinoenstatite, ferrosilite, and diopside for the silicate components (i.e., olivine and pyroxene); alabandite and oldhamite for the subgrains in the olivine; alpha iron, cohenite, $Fe_7C_3$, and magnetite for the Fe-bearing subgrains in the spinel; and Mg-Al spinel for the spinel. We used Calidris CRISP software in conjunction with the phase identification function PhIDO to index our SAED patterns of monocrystalline regions of F2-8. For polycrystalline and nanocrystalline SAED and FFT-derived diffractogram patterns, we collected d-spacing values using Gatan Digital Micrograph® DiffTools. Gatan DigitalMicrograph® was used for image processing, including diffraction patterns.

On the Nion, scanning transmission electron microscope (STEM) high angle annular dark field (HAADF) images were collected with an inner half angle of 33 mrad. A windowless 0.7 sr Bruker Xflash® SDD detector was used to collect EDS maps at a nominal current of 65 pA. EDS maps were collected for 15–20 mins with a resolution of 4 nm/pixel and over 290–591 frames. For processing of extracted EDS spectra, we used the Bruker ESPRIT software to calibrate spectra, correct the background, and check the deconvolution. Quantification was done using the Cliff-Lorimer method with detector-specific computed k factors. We excluded any thickness correction for the Cliff-Lorimer quantification given the unknown densities of some of the phases. The following elements were included in the quantification only for deconvolution purposes, as they represent contamination from sample preparation or components within the TEM or holder: Cu, Pt, Ga, Cs, and Zr. We noted excess O in all EDS analyses, which was likely due to a combination of overlapping phases in the thickness of the section, incipient hydration of the presolar grains, and oxidation of $Ga^+$ irradiated FIB sections surfaces. Owing to the difficulty in determining the relative contributions of those effects, we opted to use atomic ratios for determining stoichiometry and comparing compositional information. All EDS raw data are summarized in Table S2. EELS analyses of subgrains were performed at 200 kV with a convergence semi-angle of 30 mrad using a Gatan Enfinium ER EELS spectrometer and a 2 mm aperture with a 12.5 mrad collection semi-angle. The EELS spectra were collected for the O K core loss edge (532 eV) with a 0.05 eV dispersion and exposures of 2 s. The energy resolution for these conditions, as determined by the FWHM of the ZLP was ~ 0.4 eV. All EELS analyses were performed in Gatan Digital Micrograph®.

## 3. RESULTS

**3.1 Presolar Composite Grain and Matrix Materials**

Presolar grain F2-8 was found about 100 μm from the FIB section C89 studied by Dobrică and Brearley (2020). It has a group 1 O-isotopic composition ($^{17}$O-rich, but ~Solar $^{18}$O/$^{16}$O), as well as excesses in $^{25}$Mg, $^{26}$Mg, $^{29}$Si, and $^{30}$Si (Table 1; Fig. 1). Although some isotopic heterogeneity was observed across the composite, it can all be explained by small variations in contamination by surrounding matrix material. The isotope ratios shown in Table 1 are thus based on the most extreme values observed, which are assumed to be the least contaminated. These isotope signatures point to an origin in a low-mass AGB star, most likely with a somewhat higher than Solar metallicity (Nittler et al., 1997; Hoppe et al., 2021). The NanoSIMS images as well as scanning electron microscope (SEM) EDS maps suggested that the grain is a silicate-oxide composite, as confirmed by TEM analyses. F2-8 (Fig. 2) is a large (1.39 μm × 1.07 μm) grain, predominantly composed of Mg-rich olivine, Mg-Al spinel, and Ca-rich pyroxene; each of these components will be described in more detail in subsequent sections. There is no indication that the oxide and silicate subgrains have different isotopic compositions. The fact that similar $^{25}$Mg and $^{26}$Mg excesses are observed in both the Al-rich and Al-poor subgrains suggests that extinct $^{26}$Al does not contribute substantially to the measured Mg isotope composition, but no firm limit on the initial $^{26}$Al can be set. The outlines in Figure 2a–b show the location of the isotopic anomalies overlaid on secondary electron (SE) images. NanoSIMS and SEM-EDS maps indicate the presence of Mg-rich and Al-rich regions, which correspond to the olivine and spinel, respectively. The SE images show the polished thin section after NanoSIMS O-analysis (Fig. 2a) and after subsequent Mg analysis (Fig. 2b), hence the differences in textures. Note that the isotopically anomalous material does not extend to the top Pt marker in Figure 2b, which is consistent with the outline of the presolar grain not including matrix materials in Figure 2c. The NanoSIMS only samples the topmost 50 nm of a material, whereas FIB sections extend several micrometers in depth. Unfortunately, an attempt to re-measure the FIB section directly by NanoSIMS failed (Singerling et al., 2022). Therefore, we cannot determine exactly which portions of the FIB section, at depth, contain isotopic anomalous materials. Nevertheless, the distinct appearance of two pyroxene regions from EDS mapping, which will be discussed in section 3.4, and the close association with the olivine and spinel argue for a presolar origin for the pyroxene as well. The isotopic and TEM data for F2-8 are summarized in Table 1.

The amorphous silicate material to the right of the grain in Figure 2c–d, hereafter a-sil-A, is unlikely to be presolar, given that NanoSIMS mapping of this region does not show isotopic anomalies. Still, the a-sil-A warrants consideration owing to its close association to the presolar grain. The a-sil-A appears distinct from other amorphous silicate matrix material further down in the FIB section, hereafter a-sil-B. The two amorphous silicate materials are separated by a fracture (Fig. 2d). The a-sil-B has a higher abundance of Fe carbides and sulfides and is less continuous in its texture (Fig. 2c–d). Additionally, the appearance of the Fe carbides and sulfides embedded in the amorphous silicates of both types differ. In the a-sil-B, the Fe carbides contain magnetite rims, whereas in the a-sil-A, such rims are absent (Fig. 2d). Magnetite rims on Fe carbides have been observed in Semarkona previously (Keller, 1998; Dobrică and Brearley, 2020) and have been attributed to oxidation during aqueous alteration on the parent body. The lack of rims in the a-sil-A implies that this region has experienced less oxidation than the a-sil-B as well as the bulk of Semarkona matrix materials. Compositional data on the amorphous silicates are presented in Figure 2e, Table 2, and Table S1. The (Ca+Mg+Fe)/Si ratio and Mg#'s for a-sil-A are similar to analyses of amorphous silicates in Semarkona's matrix by Dobrică and Brearley (2020); however, a-sil-B has a higher (Ca+Mg+Fe)/Si ratio than any of the other analyses of amorphous silicates in Semarkona and also has a higher Mg# than most of the other analyses.

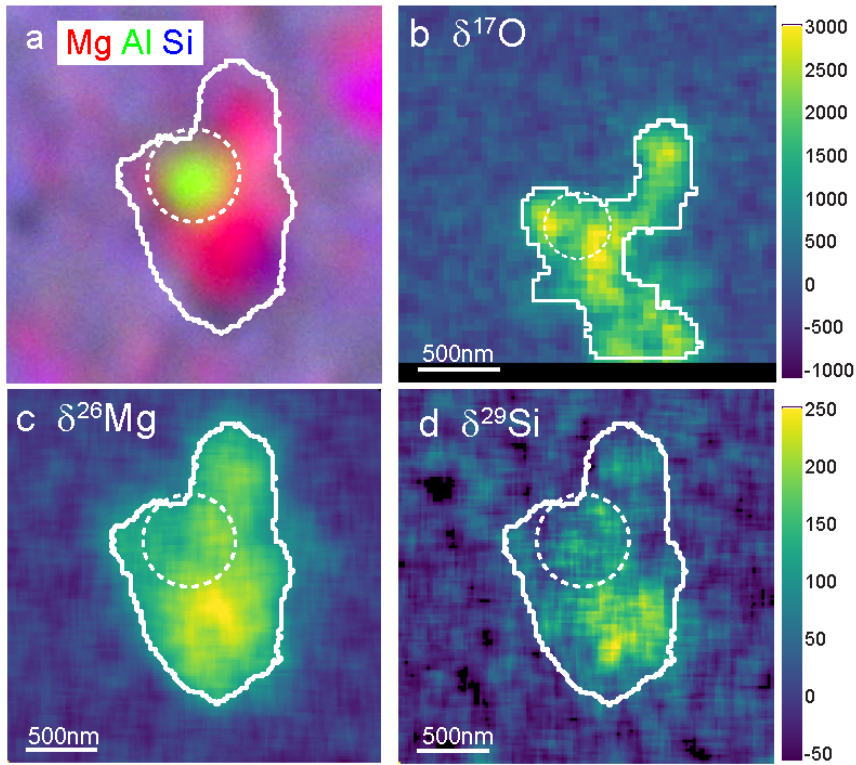

**Figure 1.** Selected NanoSIMS elemental and isotopic maps of grain F2-8: (a) Composite RGB image (red = $^{24}$Mg, green = $^{27}$Al, blue = $^{28}$Si), (b) $\delta^{17}$O, (c) $\delta^{26}$Mg, and (d) $\delta^{29}$Si isotope ratio maps. Note both that the NanoSIMS spatial resolution is higher for the O-isotope map compared to the other maps, and that the grain was identified at the edge of the automated O image, but re-centered for subsequent measurements. The dotted circle indicates the location of the spinel component. See Table 1 for the definitions of the delta values.

**Table 1.** NanoSIMS and TEM data on presolar composite grain F2-8

| | | | **NanoSIMS Data** | | | |
|---|---|---|---|---|---|---|
| Group | $^{17}O/^{16}O$ (×10$^{-4}$)[a] | $^{18}O/^{16}O$ (×10$^{-3}$)[a] | $\delta^{25}Mg/^{24}Mg$[b] | $\delta^{26}Mg/^{24}Mg$ | $\delta^{29}Si/^{28}Si$ | $\delta^{30}Si/^{28}Si$ |
| 1 | 12.09 ± 0.23 | 2.07 ± 0.03 | 211 ± 4 | 221 ± 4 | 179 ± 14 | 102 ± 17 |

| | | | **TEM Data** | | | |
|---|---|---|---|---|---|---|
| | Phase(s) | Domains | Size (nm) | Morphology | Crystallinity | Subgrains |
| **F2-8** | AOA-like | N/A | 1390×1070 | Amoeboidal | N/A | N/A |
| **Olivine** | Forsterite | 4 | 910×320, 550×170, 600×430, 510×190 | Elongate to equant | Single-xl | Oldhamite, alabandite |
| **Spinel** | Mg-Al spinel | Many | 680×320 | Elongate | Poly-xl | Magnetite |
| **Pyroxene** | Diopside-like | 2 | 530×200, 500×170 | Triangular | Weakly nano-xl | None |

[a]The O isotope values differ slightly from what was reported by Barosch et al. (2022) because that work reported the average over the whole grain, whereas we report the most extreme portion of the grain.
[b]$\delta R=1000\times(R_{meas}/R_{std} - 1)$, where $R_{meas}$ is the measured ratio and $R_{std}$ is the terrestrial ratio; N/A – not applicable, xl – crystal/crystalline.

**Table 2.** TEM EDS compositional ratios of presolar AOA-like grain F2-8's dominant components as well as amorphous silicate matrix material

| Description | (Ca+Mg+Fe)/Si | Mg#[4] | Mg/Al |
|---|---|---|---|
| Amorphous silicate A[1] | 0.80 | 45 | N/A |
| Amorphous silicate B[2] | 1.23 | 73 | N/A |
| Olivine (lo domain) | 1.96 | 100 | N/A |
| Olivine (L domain) | 1.94 | 100 | N/A |
| Olivine (R domain) | 1.90 | 100 | N/A |
| Olivine average[3] | 1.93 | 100 | N/A |
| Fe-bearing olivine (mid domain) | 1.97 | 97 | N/A |
| Mg-Al spinel | N/A | N/A | 0.45 |
| Ca-pyroxene | 0.94 | 100 | N/A |

nd – not detected, N/A – not applicable.
[1]Amorphous silicate matrix material adjacent to the presolar grain, a-sil-A.
[1]Amorphous silicate matrix material elsewhere in the FIB section, a-sil-B.
[3]Average olivine composition excludes the Fe-bearing olivine (mid domain).
[4]Mg# = (Mg/(Mg+Fe)) × 100.

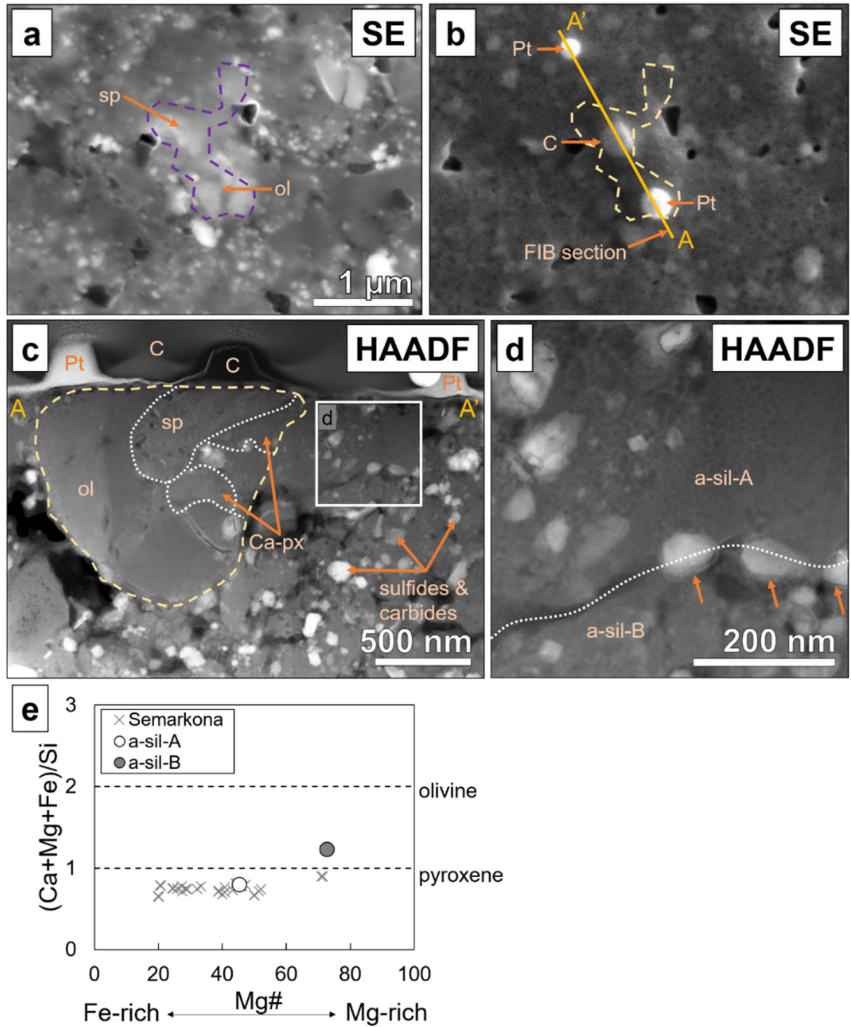

**Figure 2.** SE and STEM HAADF images and compositional data of presolar composite grain F2-8 and amorphous silicate matrix materials. (a–b) SE images show the grain as viewed on the surface of the polished thin section, with the dashed outlines indicating the extent of the grain based on NanoSIMS isotopic mapping. The SE images were collected (a) before and (b) after NanoSIMS isotopic mapping, hence the more pitted appearance in (b). (b) also shows the location of the C and Pt fiducial markers as well as the orientation of the FIB section. The HAADF image in (c) shows the overall grain along with adjacent meteoritic matrix materials viewed in the FIB section. A higher magnification HAADF image (d) illustrates the textural features of the amorphous silicates in close association with the presolar materials. The dashed white line indicates the fracture that divides the a-sil-A and a-sil-B. The fact that the a-sil-A does not have isotopic anomalies implies that these materials are solar; however, they appear to have experienced less parent body oxidation than other regions of the matrix, such as the a-sil-B, as evidenced by the absence of magnetite rims around the Fe carbides. The orange arrows in (d) point to Fe-carbide grains which show the presence of rims below and the absence of rims above the dashed white line. (e) shows compositional data comparing the two types of amorphous silicates in the FIB section to amorphous silicates elsewhere in Semarkona, as determined by Dobrică and Brearley (2020). Ol – olivine, sp – spinel, Ca-px – Ca-rich pyroxene, a-sil-A or -B – amorphous silicates.

## 3.2 Olivine Component

The olivine ($Mg_2SiO_4$) in F2-8 consists of four single-crystal domains, which are labeled as L, R, lo, and mid in Figure 3a. The morphologies and sizes of each domain are elongate and 910 nm × 320 nm for L, elongate and 550 nm × 170 nm for R, equant and 600 nm × 430 nm for lo, and elongate and 510 nm × 190 nm for mid. The four domains have similar crystallographic orientations and share multiple planes but have slight misorientations (Fig. S1). In some orientations, modulations were observed in HRTEM images of the olivine (Fig. 3b–c) and measure roughly 0.93 nm in width. The modulations are also observed as satellite reflections to the sides of some of the diffraction spots both in SAED patterns (e.g., Fig. S2) as well as diffraction patterns extracted from HRTEM images (e.g., Fig. 3c) . The fact that the satellite spots occur in non-integral spacings with respect to the substructure spots and that they tend to be diffuse implies that the modulated structure is incommensurate (Buseck and Cowley, 1983). The satellite reflections appear to be oriented parallel to (001). Additionally, the olivine was sensitive to the electron beam during initial TEM sessions, having been beam damaged during HR imaging and diffraction work (Fig. 3d). In subsequent TEM sessions, this beam sensitivity was not observed.

Compositionally, all four domains of olivine are essentially stoichiometric forsterite (Mg-rich olivine). We use the (Ca+Mg+Fe)/Si ratio as a metric for stoichiometry in silicates, with olivine having a ratio of 2, and the Mg#, calculated as (Mg/(Mg+Fe))×100, as a metric for how Mg- versus Fe-rich a silicate is, with an Mg# of 100 indicating a Mg-rich silicate. Table 2 summarizes the compositional information. The (Ca+Mg+Fe)/Si ratio of the lo, L, and R domains varies from 1.90 to 1.96, whereas the Mg# is 100. The mid domain is quite distinct from the other domains, with evidence for slight Fe enrichment (0.7 at. %) visible both in the STEM HAADF image (Fig. 3d as the brighter region) and the EDS composite X-ray map (Fig. 3e). The mid domain's (Ca+Mg+Fe)/Si ratio is 1.97 and Mg# 97. This domain also shows more voids and is adjacent to a crack that crosscuts the lo domain and is lined with Fe-rich material (Fig. 3a). With the exception of the L domain, the olivine shows elevated Al contents ranging from 0.15 at. % in the lo domain to 0.24 at. % in the mid domain and an average for the unaltered olivine (excludes the mid domain) of 0.19 at. % Al (Table S2). These concentrations are relatively high considering Al is often only present in olivine in trace amounts (tens to hundreds of ppm) (Wan et al., 2008). The EELS spectra of the olivine domains near the O K edge do not show any pre-edge features (Fig. 3f).

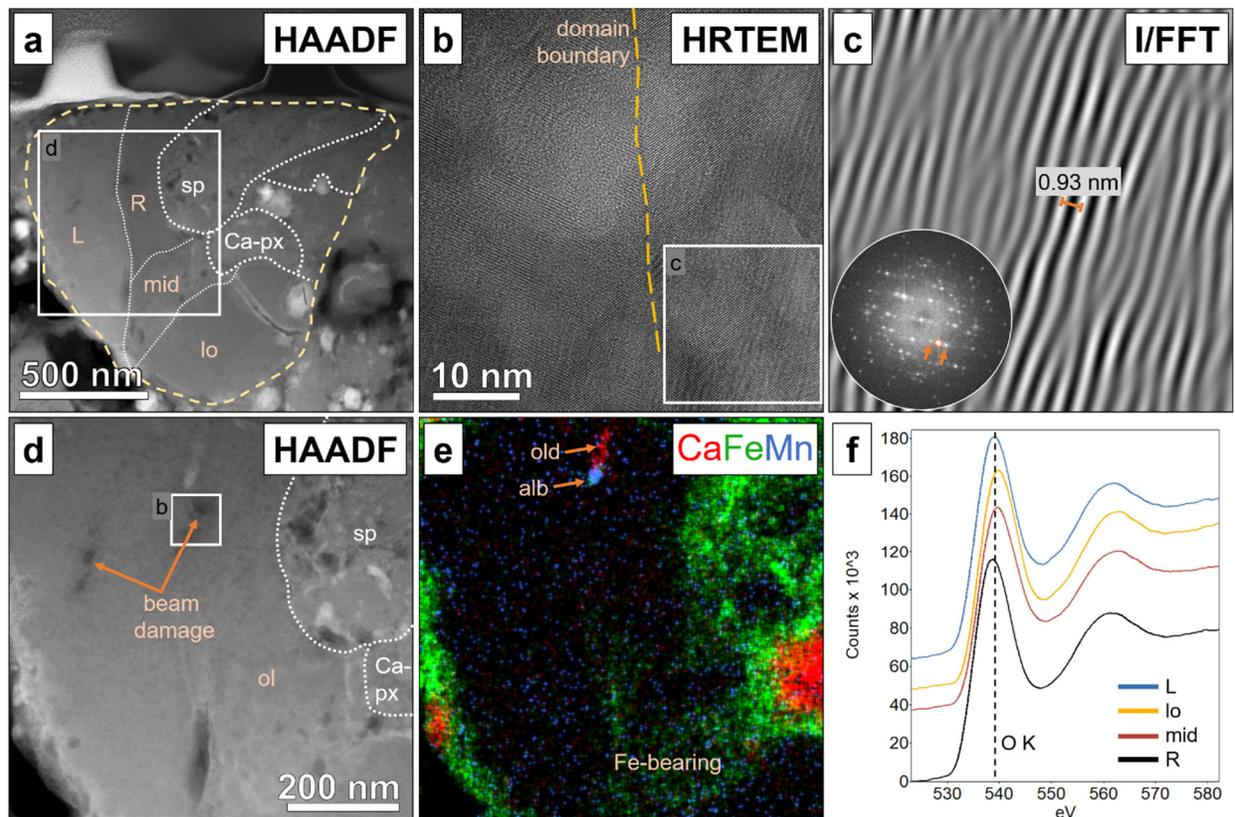

**Figure 3.** Presolar grain F2-8 olivine TEM images, diffraction data, an EDS composite X-ray map, and EELS spectra. The STEM HAADF image in (a) shows the overall grain and the four crystal domains of olivine (L, R, mid, and lo). The HRTEM image (b) shows the boundary between the L and R domains and the presence of modulations. An inverse FFT (c) and the FFT diffraction pattern it was obtained from better illustrate the modulations. The extra diffraction spots to the side of a central spot (examples indicated by arrows) are satellite reflections and come from the modulations. The inverse FFT was obtained by masking out all but the two satellite reflections around the central transmitted spot. A higher magnification HAADF image (d) shows an alabandite-oldhamite (MnS-CaS) subgrain as well as pores, two of which were the result of beam damage. The EDS composite X-ray map of Ca, Fe, and Mn (e) shows the alabandite-oldhamite subgrain in addition to Fe-bearing olivine in the mid domain. (f) The EELS spectra present data on the O K edge for the different olivine domains. The dashed vertical line corresponds to the peak rather than the edge itself, the latter of which is defined as the onset of the feature. L – left olivine domain, R – right domain, mid – middle domain, lo – lower domain, sp – spinel, Ca-px – Ca-rich pyroxene, alb – alabandite, old – oldhamite.

The only other compositional heterogeneity in the olivine is the presence of a composite subgrain at the boundary between the L and R domains (Fig. 3d–e) that is elongate and consists of alabandite (MnS), measuring 50 nm × 20 nm, and oldhamite (CaS), measuring 30 nm × 20 nm. The phase identifications were made based on EDS analyses for alabandite and oldhamite (Table 3) as well as diffraction patterns for alabandite (Table S1). The oldhamite portion of the subgrain appears to have been amorphized following EDS analyses, as there were no visible lattice fringes in HRTEM mode. Nevertheless, the identifications are consistent with sulfides rather than sulfates, as the O abundances show a marked decrease in comparison to adjacent material (-10 at.% for alabandite and -2 at.% for oldhamite, Table 3). The O abundances in sulfates (67 at.%) would actually be greater than those in olivine (57 at.%); therefore, a decrease in O abundances for both portions of the subgrain is most consistent with sulfides.

**Table 3.** TEM EDS relative compositions (atomic %) of subgrains[1] in presolar AOA-like grain F2-8, as compared to adjacent material[2]

| Description | O | Na | Mg | Al | Si | S | Ca | Mn | Fe |
|---|---|---|---|---|---|---|---|---|---|
| Alabandite | -10 | nd | -15.2 | 0.25 | 2.1 | 12.6 | 0.79 | 8.0 | 1.1 |
| Oldhamite | -2 | nd | -2 | nd | -3 | 3.3 | 4 | nd | nd |
| Magnetite 1 | 0 | 0.14 | -4.7 | -7.5 | 1.12 | 0.17 | nd | nd | 10.4 |
| Magnetite 2 | -1 | nd | -1.0 | -4.0 | 0.41 | 0.10 | nd | nd | 4.9 |
| Magnetite 3 | 0 | nd | -1.3 | -2.8 | 0.26 | 0.09 | nd | nd | 3.4 |

nd – not detected.
[1]Subgrain relative compositions were determined by subtracting the compositions of adjacent material from the subgrain original data (see Table S2 for raw data). Data less than zero are shown as negative values.
[2]Material adjacent to the subgrains includes olivine for alabandite and oldhamite and Mg-Al spinel for magnetite 1–3.

### 3.3 Spinel Component

The Mg-Al spinel ($MgAl_2O_4$) in F2-8 is an elongate, 680 nm × 320 nm, polycrystalline grain (Fig. 4a). Well-crystallized regions of the spinel have an orientation relationship with olivine, specifically the L and R domains, with shared planes only slightly offset from one another. From one orientation for which we were able to obtain diffraction patterns down zone axes for both a portion of the spinel and the L domain of olivine, we found the following relationships: $[102]_{ol}//[022]_{sp}$, $(2\bar{1}1)_{ol}//(111)_{sp}$, $(1\bar{1}\bar{1})_{ol}//(200)_{sp}$, and $(0\bar{1}3)_{ol}//(1\bar{1}\bar{1})_{sp}$ (Fig. S3). The polycrystallinity of the spinel may be a product of a secondary process, given the presence of voids and a fibrous-textured material predominantly near grain boundaries (Fig. 4c). These altered regions of spinel are enriched in Fe (Fig. 4d). In unaltered regions of the grain, the spinel is stoichiometric with an Mg/Al ratio of 0.45 (Table 2); the ideal ratio for Mg/Al spinel is 0.5.

The spinel grain also contains several Fe-bearing subgrains, which are labeled 1–3 in Figure 4c. These phases were found near altered regions of the spinel and were identified as magnetite ($Fe^{2+}Fe^{3+}_2O_4$) based on EDS (Table 3), diffraction patterns (Table S1), and EELS data (Fig. 4b). Given the small size of the subgrains and the thickness of the FIB section, it is reasonable to assume that the high Al and Mg contents in the magnetite are contributions from adjacent spinel. Comparing the subgrains to adjacent spinel, EDS spectra of the subgrains show depletions in Al and Mg and enrichments in Fe at one-to-one ratio for Al+Mg versus Fe. The O content, on the other hand, remains mostly the same. The EELS spectra of the subgrains for the O K edge is consistent with magnetite; however, all three subgrains appear to have a pre-edge feature with a peak around 533 eV. This feature is associated with hybridization of the O 2p and Fe 3d orbitals (van Aken et al., 1998). Subgrain 3 may be partly covered by adjacent Mg-Al spinel, which would explain its less obvious pre-edge feature; EELS spectra of Mg-Al spinel show no such pre-edge feature (Nyquist and Hålenius, 2014).

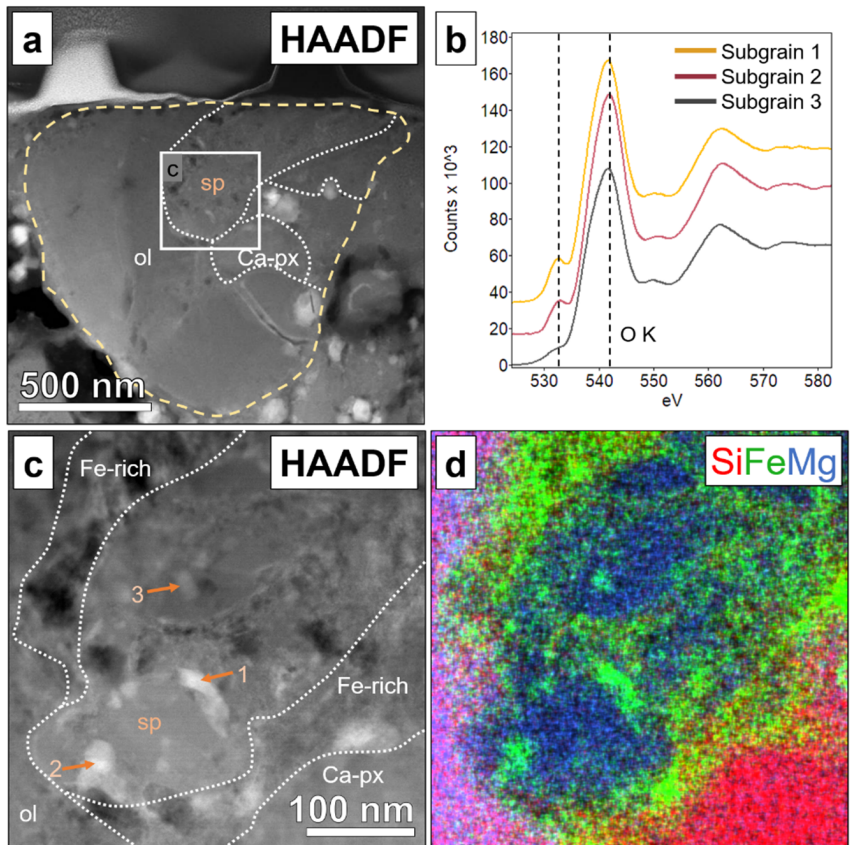

**Figure 4.** Presolar grain F2-8 spinel STEM HAADF images, an EDS composite X-ray map, and EELS spectra. The HAADF image in (a) shows the overall grain, whereas the higher magnification HAADF image in (c) illustrates the presence of magnetite subgrains (1–3) in less altered regions of the spinel. (b) The EELS spectra present data on the magnetite subgrains for the O K edge, showing a pre-edge feature with a peak of 533 eV. The dashed vertical lines correspond to peaks rather than the edges themselves, the latter of which are defined as the onset of the features. The EDS composite X-ray map of Si, Fe, and Mg (d) shows Fe-enriched, porous regions of spinel that are often correlated with grain boundaries. Ol – olivine, sp – spinel, Ca-px – Ca-rich pyroxene, Fe-rich – Fe-enriched materials.

### 3.4 Pyroxene Component

The pyroxene ($CaMgSi_2O_6$) in F2-8 consists of two triangular, 530 nm × 200 nm and 500 nm × 170 nm-sized grains separated by amorphous silicate with embedded Fe carbide, pyrrhotite, and pentlandite (Fig. 5c). The pyroxene is weakly nanocrystalline (Fig. 5b) and was sensitive to the electron beam during HRTEM imaging. Owing to the weakly nanocrystalline nature of the pyroxene, the boundaries of the grains were determined using EDS mapping, with the composite X-ray map in Figure 5d showing the two distinct regions that are enriched in Ca and depleted in Fe compared to adjacent matrix materials. Compositionally, the pyroxene is stoichiometric and diopside-like (Ca,Mg-rich pyroxene). Using the (Ca+Mg+Fe)/Si ratio as a metric for stoichiometry in silicates, with pyroxene having a ratio of 1, the pyroxene in F2-8 has a (Ca+Mg+Fe)/Si ratio of 0.94. The Mg# is 100, indicating low Fe content; however, for diopside, a more appropriate representation of its composition are the components of the pyroxene end-members, wollastonite (Wo = (Ca/(Ca+Mg+Fe))×100), enstatite (En = (Mg/(Ca+Mg+Fe))×100), and ferrosilite (Fs =

(Fe/(Ca+Mg+Fe))×100). The pyroxene in F2-8 is $Wo_{49}En_{51}$ with minor amounts of Al (Table 2). No subgrains or other heterogeneities were observed in the pyroxene.

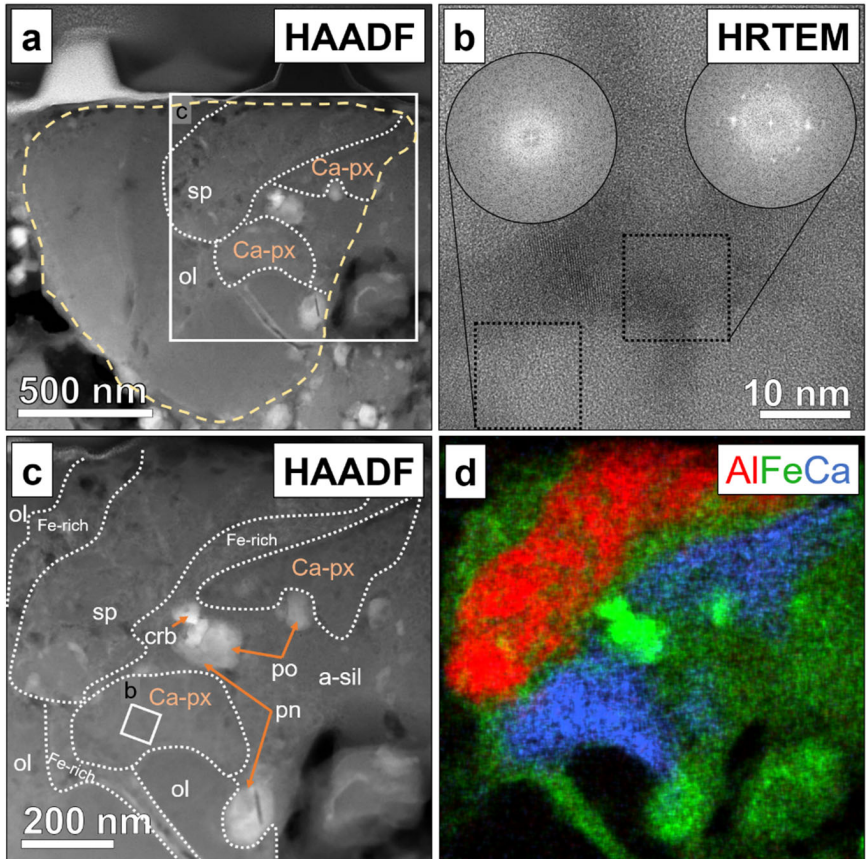

**Figure 5.** Presolar grain F2-8 pyroxene STEM HAADF and high resolution TEM images as well as an EDS composite X-ray map. The HAADF image in (a) shows the overall grain and the two domains of pyroxene. The HRTEM image (b) shows the weakly nanocrystalline nature of the pyroxene, where amorphous regions are adjacent to regions with crystallinity. A higher magnification HAADF image (c) shows the amorphous silicate material with embedded Fe carbide, pyrrhotite, and pentlandite in close association with the pyroxene. There are no obvious grain boundaries between this matrix material and the pyroxene. Instead, the EDS composite X-ray map of Al, Fe, and Ca in (d) shows the location of the two pyroxene domains. Ol – olivine, sp – spinel, Ca-px – Ca-rich pyroxene, crb – Fe carbide, po – pyrrhotite, pn – pentlandite, Fe-rich – Fe-enriched materials, a-sil – amorphous silicates.

## 4. DISCUSSION

### 4.1 Previous Composite Oxide and Silicate Presolar Grains

Previous TEM studies of presolar grains have identified oxide-silicate composite grains, all of which belong to isotopic group 1, which largely formed in AGB star envelopes although a subset has supernova origins. For example, TEM studies that have characterized presolar oxide-silicate composite grains included a spinel-non-stoichiometric silicate grain from an interplanetary dust particle (IDP) (Nguyen et al., 2014), a hibonite-non-stoichiometric silicate grain from an ungrouped C2 chondrite (Vollmer et al., 2013), and two hibonite(?)-olivine grains from CO3 chondrites (Nguyen et al., 2010; Nittler et al., 2018). The spinel-olivine grain measured 345 nm × 260 nm and had a core-rim morphology with a Mg-Al spinel core and an amorphous, Mg,Si-rich, non-stoichiometric silicate rim (Nguyen et al., 2014). The hibonite-non-stoichiometric silicate grain measured 200 nm × 150 nm with Al-rich and Ca,Si-rich regions, both of which were nanocrystalline (Vollmer et al., 2013). The hibonite(?)-olivine grains were 305 nm and 349 nm and had core-rim morphologies with Ca,Al-rich cores and Mg,Si-rich rims (Nguyen et al., 2010; Nittler et al., 2018). The grain from the Nguyen et al. (2010) study was nanocrystalline, whereas the grain from the Nittler et al. (2018) study was polycrystalline with the silicate region confirmed as forsterite (Mg# 93). In addition to the aforementioned works, Auger electron spectroscopy (AES) studies by Vollmer et al. (2009b), Floss and Stadermann (2012), Leitner et al. (2018), and Barosch et al. (2022) have observed presolar oxide-silicate composite grains as well, although AES does not allow for the same depth of microstructural and compositional characterization as TEM.

    These oxide-silicate composite grains are significant in that they represent formation under a wide range of temperatures and/or pressures. Since Al-rich oxides condense in stellar envelopes at higher temperatures than most silicates (e.g., Ebel, 2006), their co-existence requires that the oxide component remain in the envelope long enough to cool and accrete additional material, rather than being ejected away from their progenitor star by stellar winds. Additionally, oxide-silicate composites have phase assemblages reminiscent of refractory inclusions, specifically CAIs and AOAs, some of the earliest formed solids in our own Solar System that range in size from the cm- to mm-scale depending on their type (Grossman and Steele, 1976; Brearley and Jones, 1998; Krot et al., 2004; Han and Brearley, 2015). Presolar refractory inclusions can thus be compared to solar CAIs and AOAs to gain insights into similarities and differences in the dust formation of these different environments.

    Even though F2-8 is not the first oxide-silicate composite presolar grain to be identified, it is unique in its phase assemblage of spinel-olivine-pyroxene. Additionally, it has an unusual shape and is quite large (1390 nm × 1070 nm). F2-8 is also not the only irregularly-shaped presolar grain identified. Sanghani et al. (2021) observed several irregularly-shaped grains, including NWA 801_4 which may also be an oxide-silicate composite grain. Although Sanghani et al. (2022) performed TEM analyses on a subset of these grains, no observations were made on composite grains, preventing further comparisons. The irregular shape of F2-8 and the grains from Sanghani et al. (2021) could simply be a byproduct of how the grains are viewed in thin section, being potentially viewed down [111] of a cube or tetrahedra that has been bisected during thin sectioning.

## 4.2 Grain Formation in the Circumstellar Envelope

    To summarize our principal TEM observations, F2-8 contains well-crystallized and stoichiometric olivine and spinel and weakly nanocrystalline but stoichiometric pyroxene. These microstructural and chemical characterizations reflect the conditions under which the grain condensed in the envelope of an ancient AGB star. Dust condensation around AGB stars is often assumed to occur under thermodynamic equilibrium conditions, but there is evidence for non-equilibrium dust formation as well (e.g., Gail and Sedlmayr, 1999; Ferrarotti and Gail, 2001; Gail et al., 2009; Nagahara and Ozawa, 2009; Vollmer et al., 2009; Bose et al., 2012; Vollmer et al., 2013; Nguyen et al., 2016). True equilibrium condensation from a high-temperature circumstellar gas ought to produce stoichiometric and crystalline phases, with silicates tending to be Mg-rich rather than Fe-rich, since the former condense at significantly higher temperature (e.g., Sharp and Wasserburg, 1995; Lodders and Fegley, 1995; 1999; Floss and Haenecour, 2016). Non-equilibrium condensation, on the other hand, might be expected to yield non-stoichiometric and weakly to non-crystalline (amorphous) phases, with silicates tending to have more Fe than in the equilibrium case. Amorphous phases can condense when the growth of crystalline materials is

inhibited by kinetic barriers, most often during rapid cooling. In non-equilibrium condensation, kinetic effects dominate; however, in general, phases cannot form at temperatures higher than predicted by equilibrium calculations (Agundez et al., 2020), unless stabilized by surface energy or other effects not included in ideal equilibrium calculations. As such, equilibrium condensation temperatures are a useful guide for both equilibrium and non-equilibrium condensation.

Many studies have calculated equilibrium condensation temperatures for different phases but for varying pressures and C/O ratios of the progenitor stars, not all of which are applicable for the progenitor AGB star of F2-8. We present information graphically on condensation temperatures from Ebel (2006) in Figure 6, which shows condensation sequences at intermediate C/O ratios. The condensation sequences of the phases are actually more useful than the condensation temperatures themselves given that these sequences are largely independent of pressure, so long as comparisons are made for a given pressure, and allow for a broader comparison among different theoretical works. Additionally, these condensation sequences can essentially be divided into three regimes based on C/O ratios—C/O < 1.0 (M-type AGB stars, which are O-rich), C/O ~ 1.0 (S stars, which can be either O- or C-rich), and C/O > 1.0 (C stars, which are C-rich). The M stars are most consistent with the production of O-rich phases, such as oxides and silicates; whereas the C stars are most consistent with the production of C-rich phases, such as graphite and SiC. S stars are intermediate and have more complex condensation sequences (e.g., Lodders and Amari, 2005). We list the condensation sequences for these three regimes in Table 4, with the actual C/O ratio used for each regime. Data are from Ebel (2006). If we know the rough C/O ratio of the progenitor star, we can use the condensation sequences to gain insights into the histories of presolar grains. The isotopic compositional data and TEM observations from F2-8 are most consistent with C/O ratios ≤ 1.0 since dredge-up of sufficient He-shell material to substantially increase the C/O ratio would also result in large $^{26}$Mg and $^{30}$Si excesses that are not observed (Zinner et al., 2005; 2006). Note that the isotopic ratios observed in the grain are set by processes occurring before the AGB phase and thus cannot be used to precisely constrain at what evolutionary stage the grain formed.

Based on the major components in F2-8—olivine, spinel, and pyroxene—the composite grain likely represents condensation under imperfect equilibrium conditions. The olivine and spinel have characteristics consistent with equilibrium condensation, namely being stoichiometric and well-crystallized. Additionally, the olivine is Mg-rich (Mg# 100). The weakly nanocrystalline nature of the pyroxene, however, is more in line with condensation farther from true equilibrium. Previous TEM studies of presolar grains have found that olivine is more common than pyroxene and that olivine tends to be crystalline, whereas pyroxene tends to be amorphous (Vollmer et al., 2009a; Nguyen et al., 2016). This tendency for pyroxene to be amorphous has been explained by the fact that pyroxene is predicted to form from the reaction of olivine with gaseous SiO, which must happen at temperatures lower than the condensation temperature of olivine, where non-equilibrium conditions may be more common. However, this only applies to Mg-pyroxene (enstatite) that forms from Mg-olivine (forsterite). In fact, as Figure 6 and Table 4 show, Ca-rich pyroxene actually forms at higher temperatures than the condensation temperature of olivine, for most C/O ratios. As mentioned previously, kinetics allows for the formation of phases at lower condensation temperatures, which is entirely consistent with our observations. An alternative explanation for the weakly nanocrystalline structure of the pyroxene in F2-8 is that it is the product of a secondary process either in the ISM or Solar System, which will be discussed in sections 4.3 and 4.4, respectively.

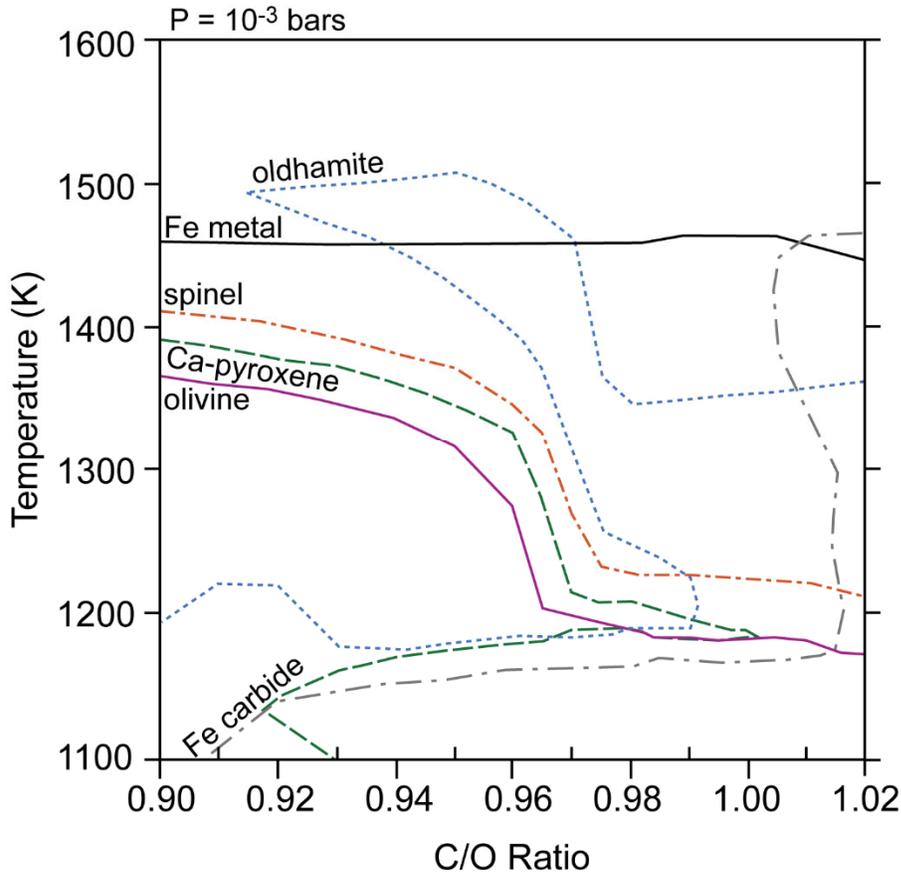

**Figure 6.** Equilibrium condensation plot showing condensation temperatures of circumstellar phases pertinent to F2-8 for a solar composition gas as a function of C/O ratio of the progenitor star. The stability fields are from Ebel (2006) and are for a static pressure ($10^{-3}$ bars). We exclude phases not discussed in relation to F2-8 from the plot in order to reduce clutter. Please see Ebel (2006) for all stability fields. Note that several phases (i.e., magnetite, alabandite) are not predicted to condense either in this range of C/O ratios and/or for these temperatures. For predicted condensates over a wider range of C/O ratios, see Table 4.

**Table 4.** Condensation sequence for circumstellar phases observed in F2-8

|  | M stars | S stars | C stars |
|---|---|---|---|
|  | C/O < 0.9 | C/O = 0.96 | C/O > 1.1 |
| Higher T | Fe metal | Oldhamite | Fe carbide |
| ↑ | Spinel | Fe metal | Fe metal |
| \| | Ca-rich pyroxene | Spinel | Oldhamite |
| \| | Mg-rich olivine | Ca-rich pyroxene | Spinel |
| ↓ | Oldhamite | Mg-rich olivine | Mg-rich olivine |
| Lower T |  | Fe carbide |  |

Data are from Ebel (2006) and are based on a static P of $10^{-3}$ bars. According to this study, Fe carbide is not predicted to form around M stars, and Ca-rich pyroxene is not predicted to form around C stars.

The elevated Al contents of the olivine (0.15–0.24 at.% Al) may provide additional insights into formation conditions. Although the substitution mechanism for Al in olivine is not clear, it likely involves coupled substitution of one $Al^{3+}$ replacing one octahedral $Mg^{2+}$ and one $Al^{3+}$ replacing one tetrahedral $Si^{4+}$, such that $MgMgSiO_4$ becomes $MgAlAlO_4$ (Grant and Wood, 2010). This substitution would correspond to percent differences in crystal radii of 24% between Mg and Al in the octahedral site and 28% between Si and Al in the tetrahedral site; for comparison, the percent difference between Mg and Fe in the octahedral site is 7%, which is consistent with the fact that the two readily substitute for one another (Shannon, 1976) (Table S3). These large differences in crystal radii could be a possible explanation for the modulations observed in the olivine (Fig. 3b–c). In terms of formation conditions, experimental work has shown that higher T and/or P can enable more Al to be incorporated into olivine's structure (Agee and Walker, 1990). Relatively high Al concentrations in meteoritic "refractory" olivines have been observed previously and have been attributed to formation by gas phase condensation in the solar nebula under oxidizing conditions (Weinbruch et al., 2000). However, these "refractory" olivines also contain elevated Ca, whereas the presolar olivine in F2-8 has Ca concentrations below detection limit. It is not entirely clear why Al but not Ca would be incorporated unless the circumstellar gas was depleted in Ca with respect to the nebular gas that formed the "refractory" olivines. An alternative explanation for the elevated Al contents in the olivine involves mobilization of Al from the spinel into adjacent phases during aqueous alteration, which will be discussed in section 4.4.

In addition to the major components of F2-8, there are several other phases present as subgrains. Given that these subgrains are located within presolar components, we ought to consider that the subgrains themselves could be presolar. The composite alabandite-oldhamite subgrain observed within the olivine represents the first observation of alabandite in any presolar grain and the first observation of oldhamite in an olivine. Oldhamite has only previously been identified in two presolar SiC grains, one which likely formed around a J star and another around an AGB star (Hynes, 2010; Singerling et al., 2021), and one presolar enstatite grain, which likely formed around an AGB star (Sanghani et al., 2022). Oldhamite is predicted to condense in circumstellar environments, but data in the literature concerning alabandite's condensation behavior are limited. Lodders and Fegley (1999) include alabandite as a condensate expected in C stars; however, the Mn-bearing phase predicted for M stars is rhodonite ($Mn_2SiO_4$) within olivine (their Table 1). Laboratory IR spectra of alabandite and oldhamite show features near 47.6 µm and 40 µm, respectively (Nuth et al., 1985; Brusentsova et al., 2012). Our identification of alabandite and oldhamite in presolar material is especially interesting because neither phase has been observed in AGB star envelopes in astronomical observational work. This could imply that these phases predominantly occur as subgrains, with their IR spectral features being obscured by their host phase, or occur in such low abundances that they are indiscernible from the spectral backgrounds.

There are three possible formation mechanisms for the alabandite-oldhamite subgrain: exsolution from the olivine host, aqueous alteration, and gas-phase condensation. We can discount solid-state exsolution for the formation of the alabandite-oldhamite subgrain given that this would require sulfur to have condensed into olivine and subsequently form a solid solution. This is not likely taking thermodynamics into account. We can also rule out aqueous alteration on the Semarkona parent asteroid (see also Section 4.4) given that Fe infiltration into the olivine does not appear to reach the part of the grain near the alabandite-oldhamite subgrain. Additionally, the close association of alabandite and oldhamite is texturally inconsistent with precipitation from a fluid. Instead, condensation is the most likely means by which the composite sulfide grain formed. The fact that there is only one rather large subgrain of this type implies that the alabandite-oldhamite grain formed first and either acted as a nucleation site or else was incorporated into the olivine as it grew. The latter seems most consistent with the fact that the subgrain is present at a boundary between two domains of olivine. Subgrains which form by condensation must form at temperatures greater than or equal to that of their host phase in order to be incorporated. Oldhamite can condense at higher temperatures than olivine, but the condensation temperature for alabandite is unknown. Condensation calculations that include alabandite are required to better constrain its actual formation temperatures.

We also identified subgrains of magnetite within the Mg-Al spinel. Presolar magnetite has been observed previously, both as a presolar grain from an AGB star (Zega et al., 2015) and as a ~30 nm subgrain in presolar graphite (Croat et al., 2008), the latter grain having likely formed in the ejecta of a core-collapse supernova. For the magnetite, there are again three possible formation mechanisms: gas-phase condensation, exsolution, and product of aqueous alteration. Pure gas-phase condensation of magnetite is not predicted as it requires oxygen fugacities several orders of magnitude above the values obtained in AGB star envelopes (Palme and Fegley, 1990). Instead, Fe metal and/or FeO condensates could have been oxidized in the AGB star envelope. Although FeO is a predicted condensate from both theoretical calculations (Agundez et al., 2020) and observational work (Decin et al., 2010), the expected oxygen fugacity is again inconsistent with AGB star envelopes (Hong and Fegley, 1998). Iron metal, on the other hand, is a predicted condensate, has been observed previously as a subgrain in presolar grains, and is not inconsistent with conditions in AGB star envelopes (e.g., Lodders and Fegley, 1999; Croat et al., 2003; Ebel, 2006; Lodders, 2006; Singerling et al., 2021). However, to explain the magnetite subgrains in F2-8, this formation mechanism would require that the subgrains condensed as Fe metal, were oxidized, and then incorporated into the Mg-Al spinel. The oxidization of metal to form magnetite would again require oxygen fugacities much greater than those expected for AGB star envelopes.

Formation of the magnetite subgrains by exsolution is another possibility. Solid solutions of magnetite and Mg-Al spinel are predicted based on thermodynamics (Nell and Wood, 1989) and would result in exsolution on cooling. There is also precedence for spinel within spinel for presolar grains. Zega et al. (2014) observed subgrains in Fe-Cr spinels that appeared to be spinel of a different composition. In that work, they argued for formation of the subgrains by exsolution owing to orientation relationships, similarities in compositions, and a correlation between subgrain size and location in the host. The magnetite for which we collected diffraction data (subgrain 3) showed a nearly identical orientation to the host Mg-Al spinel, with a euhedral morphology. There is no obvious distribution regarding size and location in the host, but the alteration of the Mg-Al spinel is largely obscuring any such relationships. The most significant argument against the formation of the magnetite subgrains by exsolution is related to the presence of $Fe^{3+}$. In the Fe-Cr spinels and their subgrains observed by Zega et al. (2014), the Fe is present as $Fe^{2+}$, consistent with expected oxygen fugacity conditions in circumstellar envelopes; however, for magnetite ($Fe^{2+}Fe_2^{3+}O_4$) to form, $Fe^{3+}$ is required. It is difficult to envision a scenario in which Mg-Al spinel that formed in an AGB star envelope was able to incorporate enough $Fe^{3+}$ to subsequently exsolve magnetite. Instead, we argue that aqueous alteration is the most probable origin for the subgrains, which will be discussed in section 4.4.

**4.3 Interstellar Medium Processing**

After formation around its progenitor star, F2-8 resided in the ISM where it was potentially exposed to a number of processes that could have altered its primary features. Two dominant processes in the ISM that may explain features observed in F2-8 include ion irradiation and grain-to-grain collisions (e.g., Klessen and Glover, 2016). Both processes tend to amorphize crystalline materials, especially sensitive phases such as the silicates. Ion irradiation ($H^+$, $He^+$, and other positive ions) from cosmic rays as well as grain-to-grain collisions have been invoked to explain the dearth of crystalline silicates observed in the ISM, as compared to those in circumstellar envelopes; crystalline silicates make up 10–20% of circumstellar silicates, whereas <2% of silicates in the ISM are crystalline (Kemper et al., 2004; 2005). However, the amorphization of materials is highly dependent on the process (ion irradiation versus shock), the energy (low or high energy cosmic rays and collisions), and the phase (spinel versus olivine versus pyroxene).

Based on experimental work, amorphization by ion irradiation happens by displacement cascades when the fluence is high enough to prevent annealing from occurring (Wang et al., 2000). Mg-Al spinel is very resistant to amorphization. For example, Wang et al. (1999) did not detect amorphization of spinel even at a fluence of $1 \times 10^{20}$ ions/m$^2$ using 1.5 MeV Kr$^+$; for comparison, olivine amorphized at $<3 \times 10^{18}$ ions/m$^2$ in their experiments. In fact, Mg-rich olivine, Mg-rich pyroxene (enstatite), and Ca-rich pyroxene (diopside) all amorphize at similar rates—$10^{18}$ ions/cm$^2$ from Demyk et al. (2004)—and become vesiculated with increasing fluence (Demyk et al., 2001; Carrez et al., 2002; Laczniak et al., 2021).

The weak nanocrystallinity of the Ca-rich pyroxene might constitute evidence for ion irradiation in F2-8, but we think that this is unlikely. Given the similar amorphization rates of olivine and pyroxene from experimental work, the two phases should show similar degrees of crystallinity. If the Ca-rich pyroxene had been nearly completely amorphized by ion irradiation, the olivine should exhibit at least partial amorphization, but that is not the case. Additionally, F2-8 does not show vesiculated textures, other than those vesicles produced from the NanoSIMS O$^-$ ion beam at the very top of the FIB section (see Singerling et al., 2022 for a more detailed explanation of this effect). Taken together, the different degrees of amorphization and the lack of vesicles in the silicates imply that F2-8 likely did not experience extensive ion irradiation.

The same argument can be made for shock from grain-to-grain collisions. Pulsed laser irradiation experiments, used as a proxy for grain-to-grain collisions, have found that olivine and pyroxene are both susceptible to amorphization from shock, with olivine actually being more sensitive than pyroxene (Weber et al., 2020). Again, if pyroxene has been nearly completely amorphized, olivine ought to show equal if not greater degrees of amorphization, which is not the case. The modulations present in the olivine are unlikely to be a product of shock given that shock metamorphism tends to instead form dislocations and planar micro-cracks in olivine (e.g., Leroux, 2001). We did not observe any of these features in the olivine in F2-8.

**4.4 Parent Body Processing**

Presolar grain F2-8 eventually became part of the molecular cloud that went on to form our own Solar System. It survived nebular processing and was incorporated into the parent asteroid from which Semarkona originated. The most common secondary processes that occurred on the asteroidal parent bodies of the ordinary chondrites include shock metamorphism from impacts, thermal metamorphism from heating produced by the decay of radioactive isotopes, and aqueous alteration from interaction of fluids originally present as ices that accreted on the parent bodies (e.g., Brearley and Jones, 1998). Semarkona is a weakly shocked sample, classified as an S2 (corresponding to 5–10 GPa; Stoffler et al. 1991), and was not exposed to elevated temperatures, as indicated by its 3.00 petrologic type. However, there are features that demonstrate Semarkona did experience aqueous alteration (e.g., Hutchison et al., 1987; Alexander et al., 1989; Krot et al., 1997; Grossman et al., 2000; 2002; Grossman and Brearley, 2005). Although the region of matrix in which F2-8 was found is very primitive, the matrix likely experienced incipient aqueous alteration. For parent body processing, we distinguish aqueous alteration that occurred at low temperatures (<200°C) from fluid-assisted metamorphism that occurred at higher temperatures (~200–400°C) (Brearley, 2006; Brearley and Krot, 2013). Given the lack of evidence that Semarkona experienced temperatures >260°C (Alexander et al., 1989), especially for the primitive region of the matrix where F2-8 was found, we exclude metasomatic effects from the subsequent discussion.

Aqueous alteration is another possible explanation for the weakly nanocrystalline structure of the Ca-rich pyroxene in F2-8. The amorphization of silicates from interaction with fluids has been invoked in previous studies of meteoritic materials, mostly recently by Ohtaki et al. (2021). However, there are a number of reasons why we argue against such a formation mechanism for the Ca-rich pyroxene. First, of the dominant minerals found in chondrites, Ca-rich pyroxene is the one of the most resistant to aqueous alteration (Zolensky et al., 1993; Hanowski and Brearley, 2001), altering significantly more slowly than olivine (Eggleton, 1986). If pyroxene had been altered, olivine should have been affected to a more extensive degree. Second, alteration of a crystalline silicate would not produce a nanocrystalline structure without also changing the composition of the material. With aqueous alteration, elements are always mobilized either into or out of the altering grain. For the pyroxene in F2-8, Ca ought to have been mobilized and Fe introduced into the pyroxene, neither of which were observed. Instead, the pyroxene in F2-8 is most likely primarily a product of non-equilibrium, low temperature condensation, rather than a secondary process either in the ISM, solar nebula, or asteroidal parent body.

However, other characteristics of F2-8 likely are the result of aqueous alteration in Semarkona's parent body. The Fe-rich material in cracks and along grain boundaries is consistent with fluid infiltration. The Mg-Al spinel shows clear textural evidence for alteration in association with the Fe-rich material, where Fe-poor regions are well-crystallized and Fe-rich regions are polycrystalline with a more fibrous appearance. We argued above (Section 4.1) that the magnetite subgrains most likely formed from aqueous alteration. Most of the magnetite subgrains are adjacent to the altered regions of the host Mg-Al spinel; for those subgrains not adjacent to altered regions, it is important to point out that the FIB section represents a 2D-slice through the spinel grain. As such, subgrains which appear to be more distant from the altered regions (e.g., subgrain 3) could, in fact, be in contact with the more altered regions in a different orientation. The irregular morphology of the spinel lends additional support to this explanation.

The olivine also shows some evidence for limited aqueous alteration. There is Fe-enrichment in association with a crack that extends from the matrix and bisects a portion of the grain. Our findings are consistent with past works that demonstrate how aqueous alteration has affected the Fe content of presolar grains (Floss and Brearley, 2014; 2015). The presence of Al in olivine could also be attributed to aqueous alteration in which Al was mobilized from the spinel. Mobilization of Al during aqueous alteration has been previously attributed to Al enrichment of altered metal inclusions by nearby Al-rich glass inclusions in CM chondrite ALH 81002 (Hanowski and Brearley, 2001). In grain F2-8, the abundance of Al increases from the L domain (below detection limit) to the lo (0.15 at.%), mid (0.18 at.%), and R (0.24 at.%) domains. The fact that the L domain is the most distant from the spinel, and the lo, mid, and R domains are just adjacent to the spinel would be consistent with Al mobilization. Additional evidence for hydrated olivine comes in the form of the beam damage we observed during HR imaging and electron diffraction. It is important to point out that these features are indicative of mild or incipient aqueous alteration, given that there was not enough O exchange to destroy the anomalous O isotopic composition of the grain and its components.

Although in section 4.2 we suggested that the modulated structure of the olivine (Fig. 3b–c) was possibly due to elevated Al contents, an alternative explanation for the features is the presence of $H_2O$. Indeed, the presence of Al in olivine has not previously been correlated to modulated structures. In general, other than dislocations, long-period defects in olivine (i.e., modulations) tend to be rare (Veblen, 1985). Exceptions include modulated structures in laihunite, an oxidized Fe-rich olivine-type mineral, and lamellar defects attributed to OH$^-$-bearing point defects in hydrated olivine (Kitamura et al., 1984; Khisina et al., 2001; Khisina and Wirth, 2002). Olivine can incorporate modest amounts of water into its structure in the form of protonated cation vacancies (e.g., Martin and Donnay, 1972; Bai and Kohlstedt, 1993; Kohlstedt et al., 1996), such as the M1 sites. The (001) plane in olivine contains a high density of M1 sites, which could explain why the modulations are parallel to that plane. Additionally, the fact that the (Ca+Mg+Fe)/Si ratios of the olivine domains in F2-8 are less than the theoretical value of 2 (Table 2) could be explained by the presence of vacancies in M1 sites, which usually contain Mg in forsterite. The sizes of the modulations (0.93 nm, Fig. 3c) are also consistent with observations of hydrated terrestrial olivines (0.9 nm) by Khisina and Wirth (2002). A pre-edge peak around 528 eV in EELS spectra is sometimes attributed to the presence of water in mineral structures (Wirth, 1997), but the EELS spectra of the olivine (Fig. 3f) did not show any such feature. Still, the interpretation of this pre-edge peak has been controversial, having been observed in water-free minerals that contain transition metals (van Aken et al., 1998) and radiation-damaged complex oxides (Jiang and Spence, 2006), and so the lack of a pre-edge feature in the olivine does not mean the mineral is anhydrous.

In terms of relative effects of aqueous alteration, it is interesting to compare the behavior of olivine and spinel to that of pyroxene in grain F2-8, especially in terms of Fe-enrichment. Although we observed Fe enrichment in most of the olivine domains and in the spinel, we did not observe significant Fe enrichment in the pyroxene. This could simply be because pyroxene is more resistant to alteration than olivine (e.g., Eggleton, 1986; Zolensky et al., 1993; Hanowski and Brearley, 2001) and perhaps spinel. Additionally, the lack of Fe enrichment in the pyroxene may be related to the extent of crystallinity, or lack thereof, between the three different phases. Both the olivine and the spinel are crystalline, whereas the pyroxene is amorphous. The ordered structures of olivine and spinel may have offered better pathways for fluid infiltration.

Based on our interpretations of the different features of composite grain F2-8, the newly condensed grain would have consisted of stoichiometric crystalline Mg-rich olivine with an alabandite-oldhamite subgrain, stoichiometric crystalline Mg-Al spinel, and a stoichiometric weakly nanocrystalline Ca-rich pyroxene. The grain would have been largely unchanged from its formation to when it was newly accreted to the Semarkona parent body, given that we found no evidence for ISM processing or alteration in the solar nebula. The final appearance of the grain after aqueous alteration would include: 1) mobilization of Al from the spinel into adjacent materials, including olivine and pyroxene; 2) introduction of Fe-rich materials, resulting in the formation of the Fe-rich fracture in the olivine, the Fe-enrichment of the olivine in the mid crystal domain, the fibrous Fe-rich material in the spinel, and the magnetite subgrains in the spinel; and 3) slight hydration of the olivine causing modulations to form. Overall, however, grain F2-8 largely escaped significant overprinting by parent body processes. The fact that oxide rims on the Fe carbides are so prevalent in matrix just adjacent to F2-8 but are not observed within the amorphous material associated with F2-8 implies that, although the presolar composite grain experienced some hydration, it did not experience it to the same extent as adjacent matrix material. The relatively pristine regions of Semarkona's matrix represent invaluable samples for future research involving presolar grains.

# 5. CONCLUSIONS

Our TEM study of presolar AOA-like composite grain F2-8 illustrates the benefits of in situ analyses to obtain a comprehensive understanding of presolar grains, enabling us to better discern the records of circumstellar, interstellar, nebular, and asteroidal processing. Grain F2-8 is the first presolar olivine-spinel-pyroxene, and adds to the growing number of refractory inclusion (CAI and AOA)-like presolar grains observed with TEM. Our observations suggest that the olivine, Mg-Al spinel, Ca-rich pyroxene, and alabandite-oldhamite formed by condensation in an M-type AGB star envelope, with the Ca-rich pyroxene condensing under non-equilibrium conditions and the other phases forming under equilibrium conditions. Additionally, this study is the first to identify oldhamite within a presolar olivine and alabandite in any presolar material to date, providing further constraints to astrophysical models and comparisons to astronomical observations.

Secondary features in grain F2-8 include Fe-enrichment in the Mg-Al spinel and olivine, elevated Al contents in the olivine, magnetite subgrains in the Mg-Al spinel, and beam sensitivity and a modulated structure for the olivine, all consistent with aqueous alteration on the parent body host but with limited overprinting of the primary features. Composite grains allow us to compare relative rates and effects of secondary processes. We found that olivine and spinel show evidence of fluid infiltration, but each reacted in different ways and to different extents. Additional high resolution NanoSIMS measurements on primitive meteorites have the potential to identify more composite grains. If they are a significant portion of the presolar grain and, by extension, circumstellar dust populations, this would have important implications for how astronomers think about dust condensation. Future IR spectroscopic work to determine the spectroscopic characteristics of such composite grains would be useful.

## Acknowledgements

We thank Associate Editor Martin Lee and three anonymous reviewers for excellent feedback and suggestions which improved the manuscript. This research was supported at NRL by NASA Emerging Worlds grants 80HQTR19T0038 (RMS) and 80NSSC20K0340 (LRN) and NASA Cosmochemistry grant NNX15AD28G (AJB).

## Appendix A. Supplementary Material

The appendix includes supplementary tables S1–S4 and supplementary figures S1–S2. The supplementary tables summarize electron diffraction data (S1), raw EDS data (S2), crystal radii for cations sites in olivine (S3), and data used in Figure 1e. The supplementary figures include: (S1) electron diffraction patterns showing orientation relationships between the different domains of olivine, (S2) diffraction and crystallographic information related to the modulated structure in the olivine, and (S3) electron diffraction patterns showing orientation relationships between the different components within presolar composite grain F2-8.

# Supplementary Figures to Tracing the history of an unusual compound presolar grain from progenitor star to asteroid parent body host

S. A. Singerling[1*], L. R. Nittler[2,5], J. Barosch[2], E. Dobrică[3], A. J. Brearley[4], and R. M. Stroud[1,5]
[1]U.S. Naval Research Laboratory, Code 6366, Washington, DC 20375, USA
[2]Carnegie Institution of Washington, Washington, DC 20015, USA
[3]University of Hawai'i at Mānoa, Honolulu, HI 96822, USA
[4]University of New Mexico, Albuquerque, NM 87131, USA
[5]Arizona State University, Tempe, AZ 85287, USA

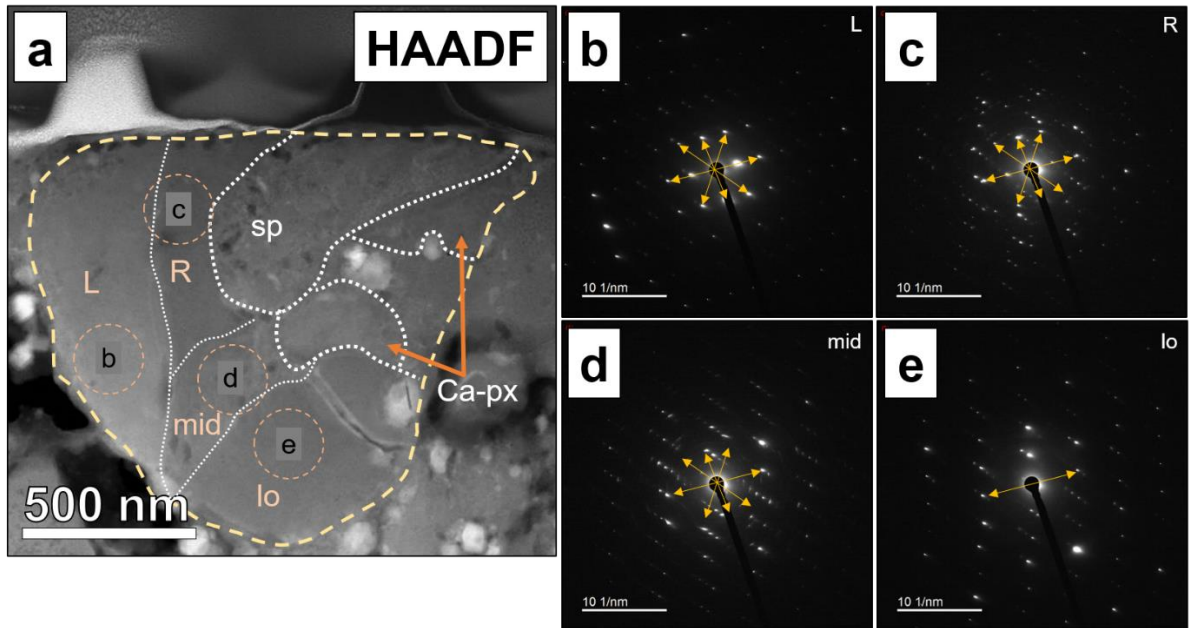

**Figure S1.** Orientation relationships between the different domains within the olivine of presolar composite grain F2-8. The STEM HAADF image (a) shows the locations where the selected area aperature was placed to collect diffraction patterns of olivine domains (b) L, (c) R, (d) mid, and (e) lo all at the same goniometer tilts. Several shared planes were observed as shown by the yellow arrows overlaid on each pattern. Sp – spinel, Ca-px – Ca-rich pyroxene.

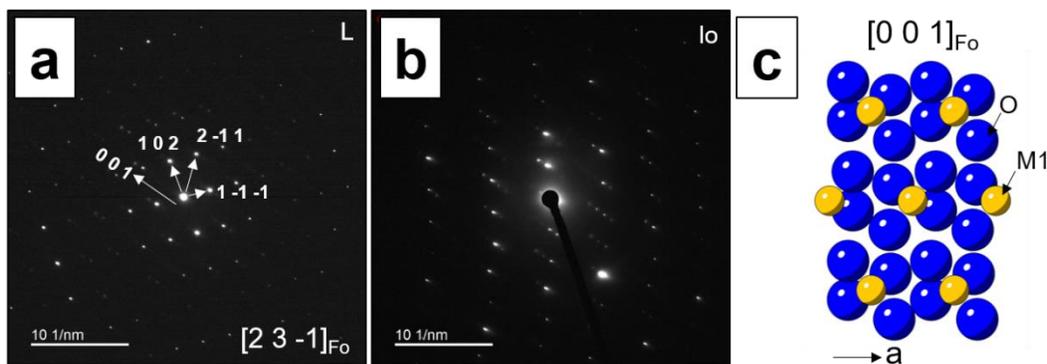

**Figure S2.** SAED patterns and crystal structure of olivine related to the modulated structure observed in F2-8. SAED patterns of (a) the L domain, indexed down the [2 3 -1] zone axis of forsterite (Fo), and (b) the lo domain, which we were not able to index owing to its curved nature. The modulated structure of the olivine is represented by the satellite reflections of the diffraction spots in the direction of (0 0 1). A model structure of forsterite as viewed down [0 0 1] shows a high density of M1 sites.

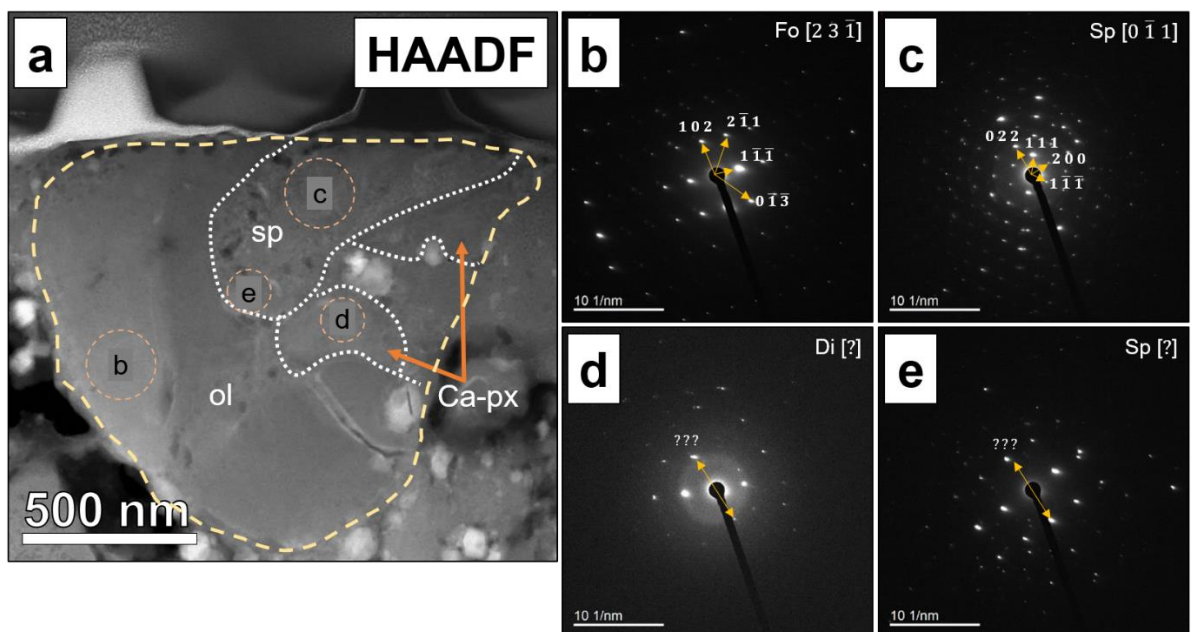

**Figure S3.** Orientation relationships between different components of presolar composite grain F2-8. The STEM HAADF image (a) shows the locations where the selected area aperature was placed to collect diffraction patterns of (b) olivine and (c) spinel at the same goniometer tilts and (d) Ca-pyroxene and (e) spinel at the same goniometer tilts. Between the olivine and spinel, the orientation relationships are: [102]ol//[022]sp, (2-11)ol//(111)sp, (1-1-1)ol//(200)sp, and (0-1-3)ol//(1-1-1)sp. A shared plane was observed between the Ca-pyroxene and spinel (yellow arrows in (d–e)), although an exact identification was not possible owing to the fact that the diffraction patterns were not collected down zone axes. The shared planes between the olivine and spinel and the Ca-pyroxene and spinel imply there is likely a genetic relationship between the phases (e.g., epitaxial growth). Ol – olivine, sp – spinel, Ca-px – Ca-rich pyroxene, Fo – forsterite, Di – diopside.

**Table S1.** Electron diffraction indexing for phases in F2-8

| Component | Description | Technique | Measured d spacing (Å) | Phase | Best Fit Theoretical hkl | d spacing (Å) |
|---|---|---|---|---|---|---|
| Olivine | L domain | SAED | 3.54 | Forsterite | 1 1 1 | 3.5027 |
| | | | 2.55 | | 0 2 2 | 2.5851 |
| | | | 2.18 | | 2 1 1 | 2.1627 |
| | | | 1.97 | | 2 3 0 | 1.9531 |
| | | | 1.77 | | 2 2 2 | 1.7513 |
| | lo domain | SAED | 3.06 | Forsterite | 1 2 1 | 3.0138 |
| | | | 1.72 | | 2 4 0 | 1.7438 |
| | | | 1.53 | | 3 1 0 | 1.5686 |
| | | | 1.38 | | 3 3 1 | 1.3996 |
| | mid domain | SAED | 3.54 | Forsterite | 1 1 1 | 3.5027 |
| | | | 3.06 | | 1 2 1 | 3.0138 |
| | | | 2.58 | | 0 2 2 | 2.5851 |
| | | | 1.79 | | 2 2 2 | 1.7513 |
| | | | 1.66 | | 2 4 1 | 1.6743 |
| | R domain | SAED | 3.60 | Forsterite | 1 1 1 | 3.5027 |
| | | | 3.06 | | 1 2 1 | 3.0138 |
| | | | 2.55 | | 0 2 2 | 2.5851 |
| | | | 2.18 | | 2 1 1 | 2.1627 |
| | | | 1.97 | | 0 4 2 | 1.9463 |
| | | | 1.79 | | 2 2 2 | 1.7513 |
| Alabandite | Subgrain | HRTEM FFT | 2.94 | Alabandite | 1 1 1 | 3.0164 |
| | | | 2.56 | | 2 0 0 | 2.6123 |
| | | | 1.86 | | 2 2 0 | 1.8471 |
| | | | 1.58 | | 3 1 1 | 1.5752 |
| | | | 1.53 | | 2 2 2 | 1.5082 |
| Pyroxene | Lower region | HRTEM FFT | 2.87 | Diopside | -3 1 1 | 2.8884 |
| | | | 2.03 | | 0 4 1 | 2.0364 |
| | | | 1.96 | | -1 3 2 | 1.966 |
| | | | 1.42 | | 5 3 1 | 1.4237 |
| | Upper region | HRTEM FFT | 2.99 | Diopside | -2 2 1 | 2.9838 |
| | | | 2.10 | | -3 3 1 | 2.1278 |
| Spinel | Unaltered | SAED | 4.75 | Mg-Al spinel | 1 1 1 | 4.6675 |
| | | | 4.17 | | 2 0 0 | 4.0422 |
| | | | 2.92 | | 2 2 0 | 2.8582 |
| | | | 2.52 | | 3 1 0 | 2.5564 |
| | | | 2.37 | | 3 1 1 | 2.4375 |
| | | HRTEM FFT | 4.56 | Mg-Al spinel | 1 1 1 | 4.6675 |
| | | | 2.84 | | 2 2 0 | 2.8582 |
| | | | 2.53 | | 3 1 1 | 2.4375 |
| | | | 2.08 | | 4 0 0 | 2.0211 |
| | | | 1.59 | | 5 1 1 | 1.5558 |
| | | | 1.42 | | 4 4 0 | 1.4291 |
| Magnetite | Subgrain 1 | HRTEM FFT | 5.03 | Magnetite | 1 1 1 | 4.8477 |
| | | | 2.82 | | 2 2 0 | 2.9686 |
| | | | 2.53 | | 3 1 1 | 2.5316 |

**Table S1 cont.**

| Component | Description | Technique | Measured d spacing (Å) | Phase | Best Fit Theoretical hkl | d spacing (Å) |
|---|---|---|---|---|---|---|
| Magnetite | Subgrain 2 | HRTEM FFT | 6.23 | Magnetite | 1 1 0 | 5.9372 |
| | | | 2.74 | | 3 1 0 | 2.6552 |
| | | | 2.36 | | 2 2 2 | 2.4239 |
| | | | 1.51 | | 4 4 0 | 1.4843 |

SAED – selected area electron diffraction, HRTEM FFT – high resolution TEM fast-fourier transform

**Table S2.** TEM EDS raw data for amorphous silicate matrix material and components of presolar grain F2-8

| Type | Description | O[1] at% | err | Na at% | err | Mg at% | err | Al at% | err | Si at% | err | S at% | err | K at% | err | Ca at% | err | Mn at% | err | Fe at% | err |
|---|---|---|---|---|---|---|---|---|---|---|---|---|---|---|---|---|---|---|---|---|---|
| Matrix | Amorphous silicate A | 71.88 | 1.93 | 2.06 | 0.21 | 4.62 | 0.34 | 2.10 | 0.23 | 12.94 | 0.51 | 0.51 | 0.13 | 0.19 | 0.10 | 0.15 | 0.09 | nd | nd | 5.56 | 0.72 |
|  | Amorphous silicate B | 69.51 | 1.89 | 1.60 | 0.18 | 10.52 | 0.57 | 1.97 | 0.22 | 11.78 | 0.48 | 0.50 | 0.13 | 0.18 | 0.09 | nd | nd | nd | nd | 3.93 | 0.57 |
| Main presolar component | Olivine (lo) | 64.28 | 1.77 | nd | nd | 23.57 | 1.04 | 0.15 | 0.07 | 12.00 | 0.42 | nd | nd | nd | nd | nd | nd | nd | nd | nd | nd |
|  | Olivine (L) | 64.61 | 1.75 | nd | nd | 23.34 | 1.01 | nd | nd | 12.05 | 0.39 | nd | nd | nd | nd | nd | nd | nd | nd | nd | nd |
|  | Olivine (R) | 63.65 | 1.73 | nd | nd | 23.68 | 1.03 | 0.24 | 0.08 | 12.44 | 0.41 | nd | nd | nd | nd | nd | nd | nd | nd | nd | nd |
|  | Unaltered olivine average | 64.18 | 1.75 | nd | nd | 23.5 | 1.03 | 0.19 | 0.07 | 12.2 | 0.41 | nd | nd | nd | nd | nd | nd | nd | nd | nd | nd |
|  | Fe-bearing olivine (mid) | 64.07 | 1.63 | nd | nd | 22.95 | 0.91 | 0.20 | 0.05 | 12.05 | 0.23 | nd | nd | nd | nd | nd | nd | nd | nd | 0.73 | 0.13 |
|  | Mg-Al spinel | 64.58 | 1.69 | nd | nd | 10.83 | 0.50 | 23.83 | 1.09 | 0.52 | 0.09 | nd | nd | nd | nd | nd | nd | nd | nd | 0.24 | 0.10 |
|  | Ca-pyroxene | 65.21 | 1.61 | nd | nd | 8.46 | 0.38 | 0.58 | 0.09 | 17.62 | 0.36 | nd | nd | nd | nd | 8.13 | 0.58 | nd | nd | nd | nd |
| Subgrain | Oldhamite | 61.55 | 3.21 | nd | nd | 21.34 | 2.07 | nd | nd | 9.97 | 1.48 | 3.27 | 0.94 | nd | nd | 3.86 | 1.23 | nd | nd | nd | nd |
|  | Alabandite | 54.37 | 1.13 | nd | nd | 7.81 | 0.28 | 0.25 | 0.05 | 15.12 | 0.18 | 12.57 | 0.55 | nd | nd | 0.79 | 0.09 | 7.96 | 0.60 | 1.07 | 0.13 |
|  | Adjacent olivine | 63.40 | 1.63 | nd | nd | 23.69 | 0.95 | nd | nd | 12.88 | 0.24 | nd | nd | nd | nd | nd | nd | nd | nd | 0.03 | 0.04 |
|  | Magnetite 1 | 64.00 | 1.49 | nd | nd | 9.84 | 0.38 | 20.03 | 0.81 | 0.93 | 0.06 | 0.10 | 0.04 | nd | nd | nd | nd | nd | nd | 5.09 | 0.45 |
|  | Magnetite 2 | 64.77 | 1.41 | 0.14 | 0.04 | 6.10 | 0.24 | 16.53 | 0.63 | 1.64 | 0.07 | 0.17 | 0.04 | nd | nd | nd | nd | nd | nd | 10.59 | 0.83 |
|  | Magnetite 3 | 64.80 | 1.55 | nd | nd | 9.54 | 0.38 | 21.19 | 0.87 | 0.78 | 0.06 | 0.09 | 0.04 | nd | nd | nd | nd | nd | nd | 3.59 | 0.34 |
|  | Adjacent spinel | 64.58 | 1.69 | nd | nd | 10.83 | 0.50 | 23.83 | 1.09 | 0.52 | 0.09 | nd | nd | nd | nd | nd | nd | nd | nd | 0.24 | 0.10 |

at% – atomic percent element, err – error, nd – not detected
Excludes elements for which all analyses were less than 0.1 at.%
[1]Oxygen excesses were from overlapping phases in the thickness of the section, incipient hydration of the presolar grains, and oxidation of Ga+ irradiated FIB sections surfaces.

**Table S3.** Crystal radii comparsions for cations sites in the olivine structure

| Element | CN | Charge | Crystal Radius (Å)[1] | Compared To: | % Diff |
|---|---|---|---|---|---|
| Mg | 6 | 2+ | 0.860 | N/A | N/A |
| Fe[2] | 6 | 2+ | 0.920 | Mg | 7 |
| Al | 6 | 3+ | 0.675 | Mg | 24 |
| Si | 4 | 4+ | 0.400 | N/A | N/A |
| Al | 4 | 3+ | 0.530 | Si | 28 |

CN – coordination number (where 6 indicates octahedral sites and 4 tetrahedral sites), % Diff – percent difference, N/A – not applicable

[1]Data from Shannon (1976)

[2]Fe is in its high spin state in olivine

**Table S4.** Data for plots

| Plot | Description | | |
|---|---|---|---|
| Figure 1e | **Series** | **(Ca+Mg+Fe)/Si** | **Mg#** |
| | a-sil-A | 0.80 | 45 |
| | a-sil-B | 1.23 | 73 |
| | Semarkona | 0.67 | 50 |
| | | 0.67 | 50 |
| | | 0.74 | 52 |
| | | 0.72 | 51 |
| | | 0.78 | 46 |
| | | 0.78 | 46 |
| | | 0.78 | 21 |
| | | 0.81 | 45 |
| | | 0.81 | 44 |
| | | 0.90 | 71 |
| | | 0.74 | 29 |
| | | 0.65 | 20 |
| | | 0.75 | 32 |
| | | 0.78 | 33 |
| | | 0.75 | 26 |
| | | 0.72 | 27 |
| | | 0.78 | 27 |
| | | 0.75 | 24 |
| | | 0.75 | 25 |
| | | 0.79 | 20 |
| | | 0.90 | 71 |
| | | 0.74 | 29 |
| | | 0.65 | 20 |
| | | 0.71 | 39 |
| | | 0.68 | 40 |
| | | 0.71 | 41 |
| | | 0.73 | 43 |
| | | 0.76 | 41 |
| | | 0.72 | 39 |
| | | 0.80 | 47 |